\documentclass[prl,aps,twocolumn,showpcs,floatfix,amsmath,amssymb]{revtex4}

\usepackage{pslatex,graphicx,dcolumn,bm,natbib}
\usepackage{bm}        
\usepackage{mathrsfs} 
\usepackage{color}

\date{\today}

\begin{document}

\title{Origin of Rigidity in Dry Granular Solids}
	\author{
	Sumantra Sarkar$^1$, Dapeng Bi$^2$, Jie Zhang$^3$, R. P. Behringer$^4$, and Bulbul Chakraborty$^1$\\
	\normalsize{
	$^1$Martin Fisher School of Physics, Brandeis University,
	Waltham, MA 02454, USA\\
	$^2$Department of Physics, Syracuse University,
	Syracuse, NY 13224, USA\\
	$^3$Institute of Natural Sciences and Department of Physics, Shanghai Jiao Tong University, Shanghai 200240 China\\
	$^4$Department of Physics, Duke University} 
}
\begin{abstract}
Solids are distinguished from fluids by their ability to resist shear. In traditional solids, the resistance to shear is associated with the emergence of broken translational symmetry as exhibited by a non-uniform density pattern. In this work, we focus on the emergence of shear-rigidity in a class of solids where this paradigm is challenged. Dry granular materials have no energetically or entropically preferred density modulations.  We show that, in contrast to  traditional solids, the emergence of shear rigidity in these  granular solids is a collective process, which is controlled solely by boundary forces, the constraints of force and torque balance,  and the positivity of the contact forces. We develop a theoretical framework  based on these constraints, which connects rigidity to broken translational symmetry in the space of forces, not positions of grains. 
We apply our theory to experimentally generated shear-jammed (SJ) states and show that these states are indeed characterized by a persistent, non-uniform density modulation in force space, which emerges at the shear-jamming transition. 
\end{abstract}
%
\maketitle

\newpage
The defining feature of a solid is its ability to resist shear.  
Persistent density modulations\cite{DFT-review}  signal the breaking of continuous translational symmetry\cite{Kurchan_Levine}, which gives rise to a non-vanishing zero-frequency shear modulus. 
This traditional paradigm of solidification is challenged by the emergence of rigidity in dry granular materials. 
Materials, such as sand,  which interact via frictional, purely repulsive contact forces have no energetically preferred structures.  Since temperature has no measurable effect on configurations of macroscopic grains, the entropic route to solidification is also unavailable \cite{ParisiZamponi_review}.  The emergence of mechanical rigidity  in these systems is a collective process controlled only by imposed stresses, the local constraints of force and torque balance, and the requirements of  positivity and friction laws on each contact force\cite{bouchaud_ideas}.  

In this letter, we develop a theory of two-dimensional (2D) granular solids  that  relates the constraints of mechanical equilibrium  to the emergence of broken translational symmetry  in force space.   
This theory  naturally  leads to a stress-only framework of rigidity  independent of the notions of strain and energy\cite{Cates1998}, and  
provides a consistent framework for understanding the properties of  experimentally generated shear-jammed (SJ) states\cite{Nature_Bi,granmat_bob}.

\begin{figure}[htbp]
\includegraphics[width=1\columnwidth]{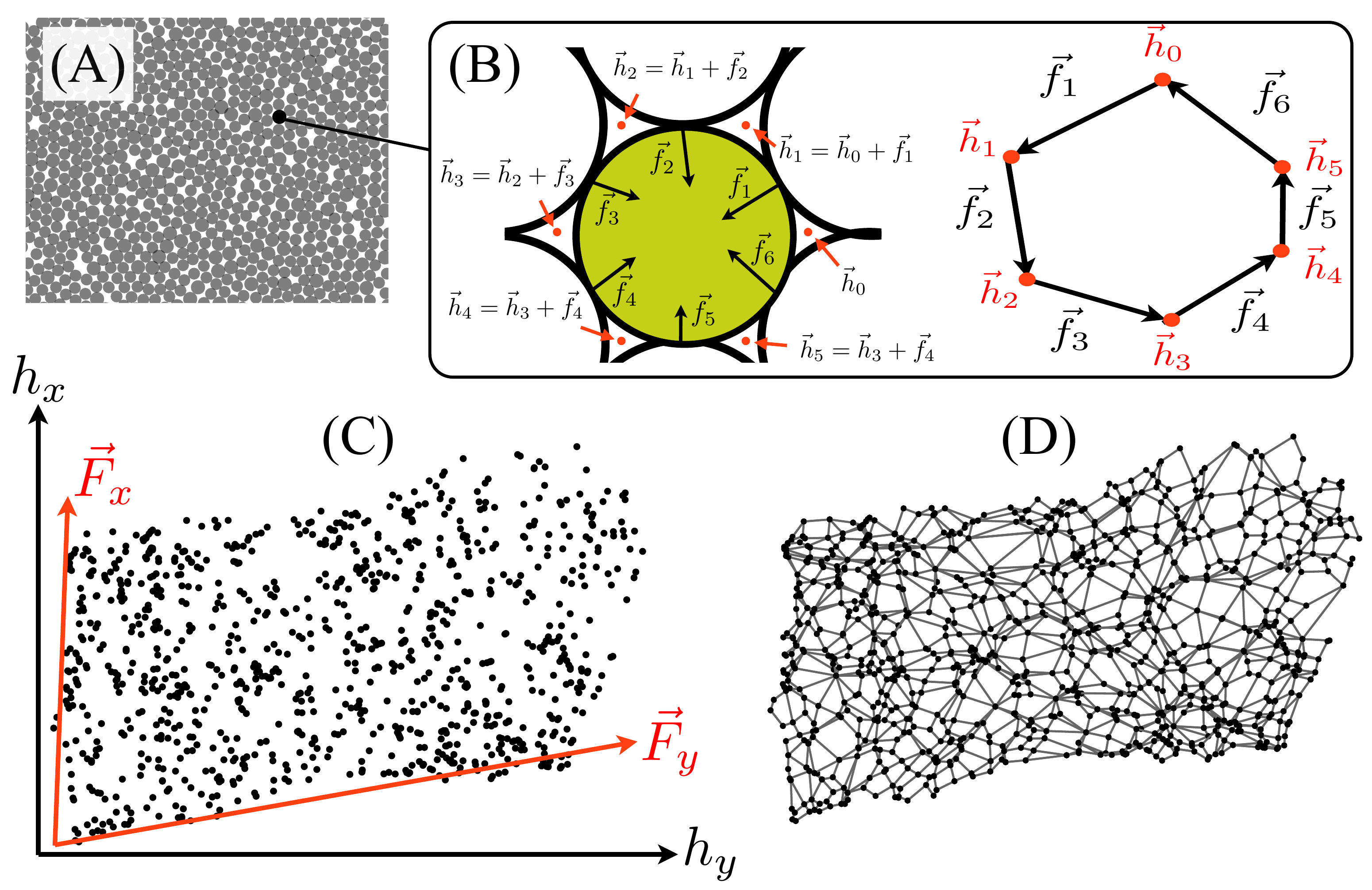}
\caption
{
(a) An SJ state in real space.
(b) 
Starting from an arbitrary origin, and going around the grain in a counterclockwise direction, the height $\vec{h}_\nu$,  is incremented by  the contact force, $\vec{f}_i$ separating two voids (red points). 
Arranged head-to-tail, adjacent forces $\vec{f}_i$,  form a force tile whose vertices are $\vec{h}_\nu$. 
(c) The height vertices of an SJ state.
(d) The force tile network constructed from the height vertices in (c). 
}
\label{forcetile}
\end{figure}

\paragraph{Mechanical Equilibrium and Height Fields}  
In a continuum formulation, the requirement of mechanical equilibrium is a divergenceless stress tensor\cite{Marder_book}, which allows for the definition of a gauge potential, $\vec{h}(\vec{r})$, such that $\hat{\sigma}=\vec{\nabla} \times \vec{h}$ \cite{Henkes,Degiuli}. There is an equivalent discrete formulation on the grain level\cite{Henkes,Degiuli}. A single-valued vector height field $\vec h$ (loop forces\cite{ball-blumenfeld}) can be defined on the dual space of the contact network or voids in a granular packing.
In this letter,  we make use of a geometric representation \cite{Tighe} (Fig.  \ref{forcetile}) that omits the real-space geometry but retains the topology of the contact network and accurately represents the structure in height-space. It can be defined as follows: the set of $\{\vec{h}\}$ values from any mechanically stable configuration form a point pattern in the height space. Since the forces on every grain add up to zero and the force associated with touching grains are equal in magnitude and opposite in direction, the point pattern in height space can be connected so it forms a polygonal tiling. The result is equivalent to the Maxwell-Cremona reciprocal tiling or force tiling (\cite{Tighe}), where each grain is represented by a polygonal tile. 

%

\paragraph{Shape of Height Space}
Under periodic boundary conditions in 2D, the force tiling in height space is contained in the parallelogram formed by two vectors $(\vec F_x, \vec F_y)$, which are related to the continuum stress field $\hat \sigma(x,y)$ by:
\begin{equation}
		\vec{F_x} = \int_0^{L_y} dy~ 
		\left[
		\begin{array}{cc}
		\sigma_{11}(x,y) \\
		\sigma_{12}(x,y)
		\end{array}
		\right]
~ ;~
 		\vec{F_y} = \int_0^{L_x} dx~ 
		\left[
		\begin{array}{cc}
		\sigma_{12}(x,y) \\
		\sigma_{22}(x,y)
		\end{array}
		\right] ~.
\label{FxFy}
\end{equation}
Here ($L_x$, $L_y$) define the size of the sample.
Since $\hat \sigma(x,y)$ is the curl of $\vec{h}$, 
$(\vec{F}_x, \vec{F_y})$ represent the difference in $\vec{h}$'s across the sample.   These vectors represent the amount of mechanical load present in the system, and are related to the force moment tensor
$\hat{\Sigma}$ via
\begin{equation} 
\hat{\Sigma} \equiv\sum_{i \neq j} \vec{r}_{ij} \otimes \vec{f}_{ij} = 
\left(\begin{array}{cc} 
L_x & 0 \\ 0 & L_y
\end{array} \right) 
\times 
\left(\begin{array}{cc} 
\vec{F}_x \cdot \hat{x}, & \vec{F}_x \cdot \hat{y} \\ 
\vec{F}_y \cdot \hat{x}, & \vec{F}_y \cdot \hat{y} 
\end{array} \right) 
\label{Fvec_FMT},
\end{equation} 
where $\vec{r}_{ij}$ is the contact vector from the center of
grain $i$  to the inter-particle contact between
grains $i$ and $j$, $\vec{f}_{ij}$ is the force vector
associated with this contact, and $ \otimes$ is the outer product.
 The definition of $(\vec F_x, \vec F_y)$ also shows that applied stress is necessary to obtain a non-zero height difference, and hence delineate the space within which the height vectors lie.  For  a finite system, 
 it is still possible to define global vectors representing height differences, however,  the box enclosing the height pattern is no longer a parallelogram.  An average parallelogram can be defined via the columns of the force-moment tensor  $\hat{\Sigma}$, as shown in  Fig. \ref{forcetile}.

If  a height pattern, as defined by the density field $\rho(\vec h) = \sum_i \delta(\vec h - \vec h_i)$, does not change with respect to small, continuous deformations of its boundary, we will define such a structure to have  persistent order in height space. 
Since changes to the boundary of the force tiling 
in height space is equivalent to changing the loading stresses on the sample, a granular assembly created at a given $(\vec F_x, \vec F_y)$, will collectively resist shear deformation,  if it has persistent order.   The question we ask is whether the constraints of mechanical equilibrium can lead to persistent order, and under what conditions.


\paragraph{Positivity, Coulomb conditions and Convexity}
Since the forces can be arbitrarily small, the heights are continuous variables and the height vertices can be any set that fits in the parallelogram defined by $\vec F_x$, $\vec F_y$.  Each such height configuration would correspond to a force-balanced granular assembly consistent with the stresses imposed at the boundary.
In the absence of any other constraints, therefore, we would not expect any correlations or broken-translational invariance in height space.  There are, however, two inequality constraints that have been left out of our analysis so far, and as we show here,  these constraints can be translated to the geometrical requirement that, statistically,  the force tiles (polygons) representing assemblies of dry grains are convex.

\begin{figure}[htbp]
\begin{center}
\includegraphics[width=1\columnwidth]{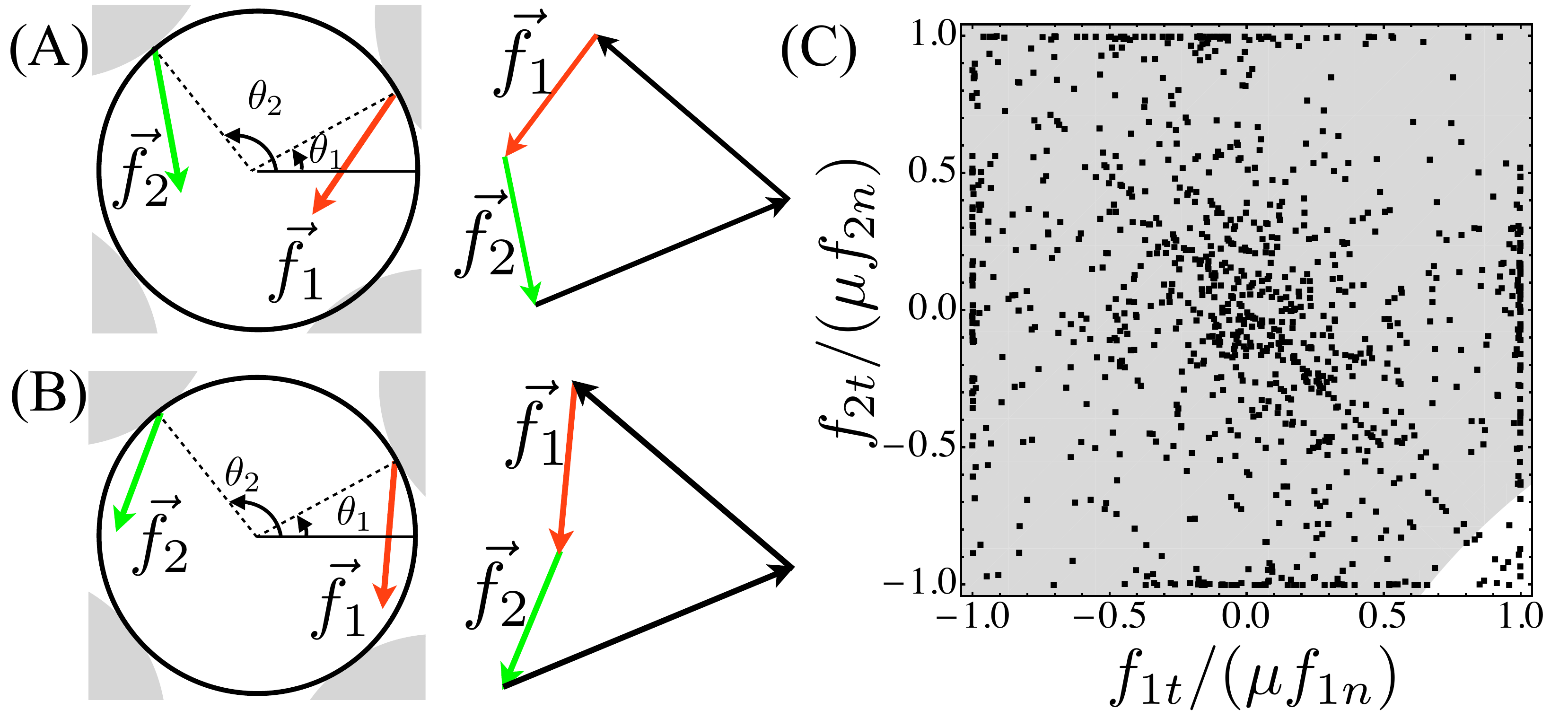}
\caption
{
Two consecutive forces can form part of a convex (A) and concave polygon (B). 
%
(C) Convexity map for $\theta_2-\theta_1 \ge \pi/3$  (Eq.~\ref{convexity_cond}) and $\mu=0.7$. Grey region denotes convex and white concave. Experimental data from a typical SJ state is also shown (points). 
}
\label{convexity}
\end{center}
\end{figure}

For a granular solid composed of circular grains and only frictionless forces between grains, all forces are central to the grain. Upon a rotation of 90$^\circ$, all forces become tangential to the grains. A convex polygon that exactly inscribes the grain can be constructed by simply elongating the rotated force vectors. This polygon is related to the force tile by a conformal transformation. Hence all force tiles are convex if the forces are frictionless. 
It is possible to have concave polygons for force tiles when frictional forces are considered. Two consecutive forces around a grain can either form a convex vertex (Fig.~\ref{convexity} A) or a concave one depending on how frictional they are and the angular distance $\theta_2-\theta_1$ between the contacts. Decomposing each force into tangential and normal parts, the condition for convexity can be easily obtained:
\begin{equation}
1+\frac{f_{1t} f_{2t}}{f_{1n} f_{2n}} +\left(\frac{f_{1t}}{ f_{1n}}-\frac{f_{2t}}{ f_{2n}}\right) \cot(\theta_2-\theta_1) \ge 0, 
\label{convexity_cond}
\end{equation}
where the tangential force and the normal force obey the Coulomb criterion for a given static friction coefficient $\mu$: $\frac{f_{1t}}{f_{2t}}, \frac{f_{1n}}{f_{2n}} \le \mu$. The angular distance between two contacts $\theta_2-\theta_1$ is constrained by geometry.   In a mono-dispersed packing of just-touching disks, for example,  $\theta_2-\theta_1$ cannot be smaller than $\pi/3$. Using this as a lower bound, Eq.~\ref{convexity_cond} gives the range of values of tangential forces for which convexity is possible for a given $\mu$. A straight forward calculation based on  Eq.~\ref{convexity_cond} shows that for any $\mu<1/\sqrt{3}\simeq 0.58$, the convexity condition is never violated. For  $\mu=0.7$ which is the static friction coefficient of the particles studied in the experiments\cite{Nature_Bi,granmat_bob}, it is possible to have concave edges when the two consecutive forces $\vec{f}_1$ and $\vec{f}_2$ are  simultaneously fully mobilized contacts (Fig.~\ref{convexity}(C)). In Fig.~\ref{convexity}(C), we also show the forces of a typical SJ state. While a few contact pairs form concave edges, they are rare occurrences and we deduce that, statistically, force tiles are convex for typical physical values of $\mu$. 

The two inequality constraints, positivity of $f_{in}$, and the Coulomb criterion, $f_{it} \le \mu f_{in}$, are the most difficult to implement in any statistical mechanics calculation\cite{Degiuli}.   The previous discussion indicates that these constrains can be effectively captured as a convexity constraint on the force tiles.
If unconstrained in height space, the ensemble of all possible point patterns formed by the vertices are trivially expected to have a liquid order or $\langle\rho(\vec h)\rangle=const$.
With the requirements of convexity and tiling in height space, the vertices of a tile cannot come arbitrarily close to each other. This requirement constrains the possible point patterns formed by the vertices of the tiles to a much smaller subset of configurations, hence giving rise to the possibility of broken translational symmetry in height space or  $\langle\rho(\vec h)\rangle \ne const$.
 The constraints act as effective springs that tie the vertices to their average positions. If these springs constraint the position of every vertex in the tile to a region that is small compared to the average force (length of a link), then we expect to see correlations and broken translational invariance in height space.  The strengths of the effective springs are not predetermined but emerge as a consequence of the local constraints and the global constraints through $\vec F_x$, $\vec F_y$, or $\hat \Sigma$.   

Based on the above points, we argue  that broken translational symmetry and persistent order emerges in height space as the number of vertices is increased through the creation of force-bearing contacts between grains as a set of grains is stressed.   In the remainder of this paper, we construct and analyze height patterns of experimentally generated SJ states, and show that rigidity is concurrent with appearance of persistent order and occurs at a critical value of the fraction of force-bearing grains\cite{Nature_Bi}.

\paragraph{Correlations and Rigidity in experimental SJ states}  
As an example of a solid created though a distinctly non-equilibrium process of imposed stresses without any changes in the density, we study the height pattern and force tiles of SJ states. 
These states are created through a quasi static process of pure shear that preserves the area at packing fractions below the minimum required for creating isotropically jammed states\cite{Nature_Bi,Ren_Dijksman}.  It has been shown earlier that the parameter that controls the transition to SJ states is the fraction of force-bearing grains, independent of the nominal packing fraction\cite{Nature_Bi}.   A hallmark of these states is that pressure increases with shear strain\cite{Nature_Bi,Ren_Dijksman}. 

\begin{figure}[htbp]
\begin{center}
\includegraphics[width=1\columnwidth]{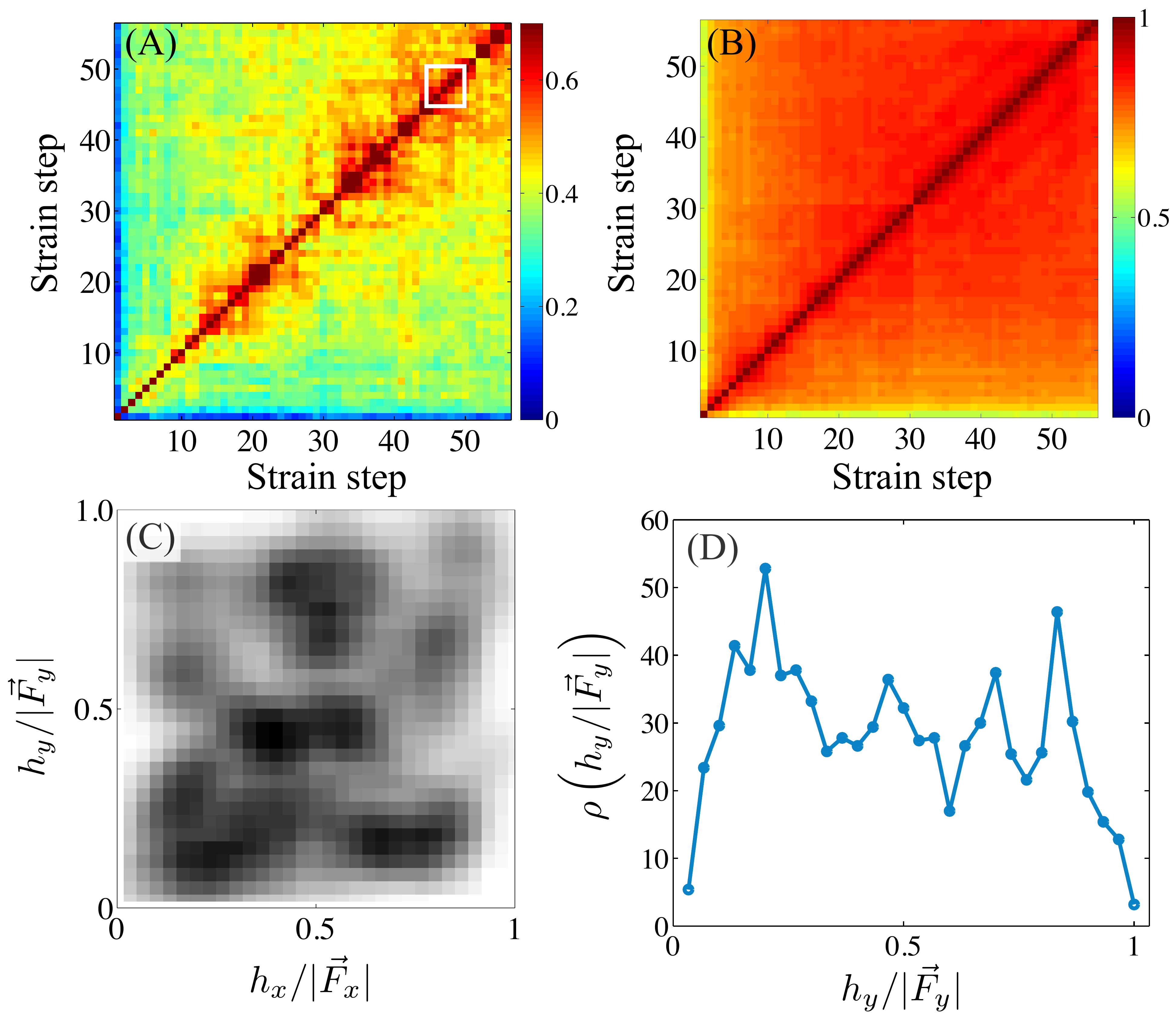}
\caption
{ (A) Color map of the overlap matrix, $d^{\alpha,\beta}$  
(B) Same as (A) but for real-space density of grains.  (C)   $\langle \rho({\vec h}) \rangle$, from states within the white box marked in (A). (D) A one dimensional projection of $\langle \rho({\vec h}) \rangle$. The bounding box in height space has been divided into $30 \times 30$ grid, on average, accommodating 5 vertices per bin.  }
\label{overlap}
\end{center}
\end{figure}

Fig.~\ref{forcetile} shows the tiling and the height point pattern for a SJ state.    In order to investigate the emergence of persistent order under the shear-jamming process, we define an overlap function between configurations $\alpha$ and $\beta$ at two different strain steps as: $d^{\alpha,\beta} = \sum_{m,n} \rho_{m,n}^{\alpha} \rho_{m,n}^{\beta}$.  The density field $\rho_{m,n}$ is defined on a rectangular grid that stretches affinely with $(\vec F_x, \vec F_y)$, and is obtained by binning the height point pattern.  If the point pattern undergoes a completely affine transformation from one strain step to another, then the overlap is unity.  Fig.~\ref{overlap} shows the overlap matrix $d^{\alpha,\beta}$ for one strain history at a packing fraction of $0.79$.   The overlap matrix clearly shows that the density patterns become persistent at the higher strain steps, and the SJ states are characterized by persistent order of the height pattern.  The persistent pattern also exhibits broken translational symmetry, as evidenced by the averaged density pattern shown in Fig. \ref{overlap}.  The average density is defined as $\langle \rho_{m,n} \rangle = \frac{1}{M} \sum_{\alpha=1,M} \rho_{m,n}^{\alpha}$, where the sum runs over strain steps within the shear-jammed regime of the strain history.  This analysis demonstrates that the SJ states created by shear can sustain further shearing: stress-only elasticity. 
It was shown (\cite{Nature_Bi}) that the onset of rigidity during the SJ transition is signaled by the percolation of the strong force network, or equivalently where the fraction of non-rattlers reaches a threshold value, $f_{NR}=0.84$. We find that the onset of order in height space coincides with the onset of rigidity as defined by $f_{NR}$.

The above analysis demonstrates that  the SJ states can sustain incremental loading because of the emergence of broken translational symmetry in the height point pattern.  The mechanical response of these solids  is determined by  the response of this pattern to stress.  This, stress-only,  description provides  an exact parallel to the framework of elasticity of thermal solids, which describes the response of the real-space density to strain.

We have analyzed the evolution of the positional density of grains, $\rho(\vec r)$ along the strain history and find negligible changes.   The grains are in nearly touching configurations, and contacts break and form during the strain history with no significant effect on any positional correlations, as seen from the overlap of real-space densities in Fig. \ref{overlap}.   We find that the  pair correlation function, $g(r)$, is typical of any disordered solid and insensitive to the SJ transition.  The signature of SJ which is a zero-temperature, stress-induced solidification  is manifested only in the height pattern\cite{Ikeda:2012fk,Teitel_arxiv2012}. 

\begin{figure}[htbp]
\begin{center}
\includegraphics[width=1\columnwidth]{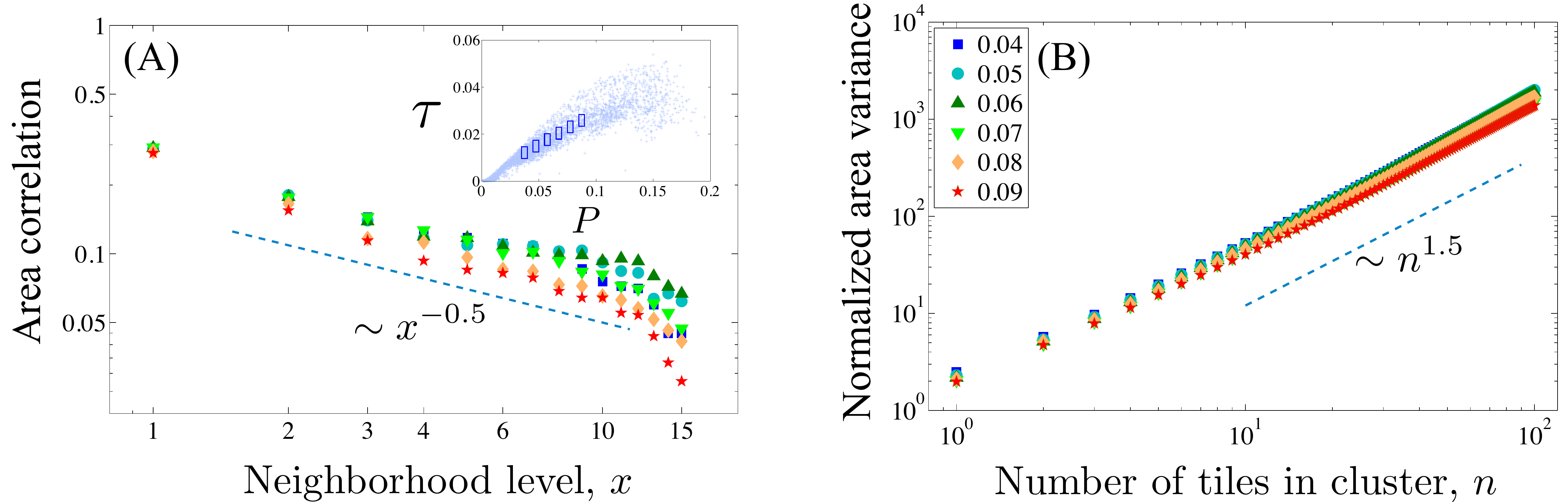}
\caption
{
(a): Connected correlation function of tile areas for  SJ states  grouped by pressure (inset).
(b):  The variance of the tile-areas as a function of the cluster size ($ n $). The exponent, 1.50, is consistent with the exponent seen in the correlation function decay . Inset (a): The ensemble of SJ states in the $ P-\tau $ phase space.  Each  blue box  is labelled by the average pressure of enclosed states. 
}
\label{correlation}
\end{center}
\end{figure}

In addition to studying $\rho(\vec h)$, we have also analyzed height-space correlation functions in the ensemble of SJ  states.  The ensemble is well characterized by the average shear stress $\tau$, and the pressure, $P$\cite{Nature_Bi,EPL_Bi}.  The packing fraction plays a less crucial role in the  the SJ states\cite{Nature_Bi,EPL_Bi}.  Fig. \ref{correlation} shows the two-point correlation of the areas of force tiles associated with two grains that are separated by a certain neighbor distance.  It should be emphasized that this is not a metric distance between grains in real space.  Grains that are separated by $x$ neighbors are not separated by a fixed distance between their positions.  In fact, the path between two such grains, via inter grain contacts,  can meander through the granular assembly because of the presence of rattlers.   The neighbor-level correlations decay as a power law with an exponential cutoff that depends only on  $\tau$ and $P$.   The decay is slowest for the states closest to the onset of shear jamming, which form a line in the zero temperature jamming phase diagram ending at  $\phi_J$\cite{Nature_Bi}.  

The implication of this area correlation is a non-trivial scaling of the variance of the determinant of the stress tensor with the number of grains, which can be measured in real space.   If correlations of the area were short ranged, the variance of the determinant of the stress tensor for $n$ grains (equivalently $n$ force tiles) should scale as $n$.   On the other hand, if the correlations fail to decay by  $\approx \sqrt{n}$ neighbors level, then their integral, which gives the variance, would scale super-extensively with $n$.  As shown in Fig. \ref{correlation},  and in Ref. \cite{EPL_Bi}, for $n$ ranging from $5$ to $100$, the variance scales as $n^{1.5}$.   The scaling of the variance can be easily measured in experiments, whereas measuring the height-space correlations require construction of the height map.  The nature of correlations in height-space can, therefore, be deduced from measurements of the variance of the stress.   It should be remarked that the variance of stress in isotropically jammed states scales as $n$\cite{Puckett_Daniels} suggesting a lack of long-range correlations in the height point pattern of these states.   The SJ states seem, therefore, to be qualitatively different from states above $\phi_J$, and it is intriguing to speculate that 
$\phi_J$ marks the transition from lack of a distinct signature of jamming  in real-space  and strong signature in the height pattern to the dual scenario of strong signature in real space\cite{Ohern_2003} and no distinctive signature in the height pattern.

\paragraph{Discussion} We have shown that the constraints of mechanical equilibrium in zero-temperature assemblies of dry grains lead to the emergence of broken translational symmetry in a space of gauge potentials\cite{Degiuli}, the height fields.  The necessary condition for persistent order is encoded in a geometrical constraint of convexity on the shape of force tiles formed by connecting the heights corresponding to a single grain. This geometrical constraint is a consequence of two inequalities: positivity of the normal forces, and the  static equilibrium restriction on the range of the tangential  forces.  Persistent order develops as more and more force bearing contacts are introduced into a grain packing, which translates to an increase in the number of height vertices.  The process is thus reminiscent of density-driven solidification, albeit in a space that refers to forces and not positions of grains.  The persistence of order in SJ states, shown in Fig. \ref{overlap}, has an interesting structure.  The range of shear strain over which a shear-jammed solid resists plastic failure is visible as the squares of warmer colors, which clearly grow in number as the shear stress creating the SJ state increases.  Fig. \ref{convexity} shows that non-convex polygons are present in the SJ states. 
We have preliminary evidence that the failing of a cluster of non-convex polygons leads to a rapid decrease of overlaps in Fig. \ref{overlap}a.   We will perform a detailed analysis of the connection between non-convex polygons and plastic failure in the near future.

The height-space picture provides a description of elastic and plastic behavior of assemblies of dry grains by referring only to their stress state specified through $\vec F_x$ and $\vec F_y$.  This stress-only description avoids any reference to the concepts of strain and energy, which are difficult to define unambiguously in assemblies of dry grains\cite{Cates1998}. 

Our analysis has been restricted to 2D.  The tiling picture does not extend to 3D and presence of gravity.  An analog of the height fields does exists in 3D\cite{Henkes,Degiuli}, and a completely parallel structure can be constructed through Delaunay triangulation of the grain network in real-space\cite{Degiuli}.   
It is, therefore, plausible that the general concept of order in height space extends to 3D, and  \cite{Degiuli} provides a mathematical framework for developing and testing a theory of rigidity in 3D.
 
\begin{acknowledgments}
This work was partially funded by NSF-DMR-0905880.  
Dapeng Bi acknowledges the support of NSF-IGERT fellowship and Syracuse University.
Jie Zhang acknowledges the support from SJTU startup fund and the award of the Chinese 1000-Plan (C) fellowship. 
R.P. Behringer acknowledges support provided by NSF-DMR-1206351 and ARO Grant No. W911NF-1--11-0110.
The authors also acknowledge useful conversations with Susan Coppersmith, Albion Lawrence, Matt Headrick, Patrick Hayden, Taylor Hughes, Mitch Mailman, Lisa Manning, and Aparna  Baskaran.

\end{acknowledgments}

\bibliographystyle{apsrev}
\bibliography{references}

\end{document}